\documentclass[aps,prl,twocolumn,superscriptaddress]{revtex4}
\usepackage{graphicx}
\usepackage{epsfig}
\usepackage{epstopdf}
\usepackage[mathscr]{eucal}
\usepackage{amsmath,amssymb,bm,graphics}
\usepackage[pdftex,letterpaper=true]{hyperref}

\renewcommand{\bar}[1]{\overline{#1}}

\renewcommand{\thefootnote}{\fnsymbol{footnote}}
\renewcommand{\bar}[1]{\overline{#1}}
\begin{document}

\setcounter{footnote}{0}
\renewcommand{\thefootnote}{\fnsymbol{footnote}}
\renewcommand{\bar}[1]{\overline{#1}}

\preprint{SLAC-PUB-13422}

\title{Light Front Holography: Response to a Comment by L. Glozman}

\author{Guy F. de T\'eramond}
\affiliation{Universidad de Costa Rica, San Jos\'e, Costa Rica}

\author{Stanley J. Brodsky}
\affiliation{SLAC National Accelerator Laboratory, Stanford University,
Stanford, California 94309, USA}

\date{\today}

\begin{abstract}

We reply to L. Ya. Glozman's   Comment ``Is a consistent holographic description of excited hadrons with fixed  
$L$  possible?"  In general, the solutions of the bound state Hamiltonian equation of motion in QCD
have Fock state components with different $L$, and consequently hadronic  relativistic  wave equations derived in
the framework of light-front holography share this property. For example, the proton eigenstate has components with $L=0$ and $L=1$. The results of AdS/QCD are consistent with the requirements of chiral symmetry and do not violate any 
fundamental physical principle. This is further illustrated with various examples.

\end{abstract}


\maketitle


We have demonstrated in our recent Letter~\cite{deTeramond:2008ht}  the equivalence of Anti-de Sitter (AdS) five-dimensional background space  representation with light-front (LF) quantized theory in physical 3+1 space-time. The result is a single-variable light-front Schr\"odinger equation which determines the eigenspectrum and the light-front 
wavefunctions of hadrons for general spin and orbital angular momentum.  In the case of the soft-wall model with massless quarks, the hadronic spectrum has the form of the Nambu string: $M^2 = 4 \kappa^2(n+ L+ S/2)$ in agreement with
experimental results~\cite{Klempt:2007cp}. The scale  $\kappa$ is determined by the soft-wall dilaton metric, $n$ is the principal quantum number, $L$ characterizes the {\it minimum } internal orbital angular momentum of the state and $S$ is the total quark spin.   The pion with $n=0, \, L=0, \, S=0$ is massless, consistent with chiral symmetry.

In the Comment~\cite{Glozman:2009fj}   L. Glozman argues that 
``it is impossible in general to construct a holographic description of 
each meson (with the quark-antiquark valence degree of 
freedom only) with fixed (conserved) quantum number $L$ 
for each meson that would satisfy at the same time 
manifest chiral symmetry in the ultraviolet and would 
not violate unitarity."

In general, the hadronic eigensolutions of the LF Hamiltonian equation of motion, and consequently of the AdS equations derived in the framework of light front holography~\cite{deTeramond:2008ht}, have components which span  a set of orbital angular momenta $L$, not a single fixed value, as stated in \cite{Glozman:2009fj}.  For example, as we have shown in \cite{Brodsky:2008pg},  the  spin-1/2  eigensolution for the proton has a Dirac two-component spinor structure 
$\Psi_\pm$ where the upper component has $L=0$ (parallel quark and proton spin) and the lower component has $L=1$ (antiparallel quark and both spins).  Both components are eigensolutions with the same hadronic mass,  consistent with chiral symmetry.  
However, the different orbital components of the hadronic eigensolution have different behavior  as one approaches the short distance $z \to 0$ boundary as predicted from the twist of the relevant interpolating operator.  Only the minimum $L$ term survives at the short distance $x^2=0$, $z=0$ boundary, in agreement with the AdS/CFT dictionary. Thus, only the $S$-wave of the proton eigenstate couples to the relevant interpolating hadronic operator  at $z \to 0$;    the $L=1$ component vanishes with one extra power of $z$.  
Indeed, the relative suppression of the proton LFWF  $\Psi_{-}$ at $z\to 0$ leads to the observation that the struck quark in the proton structure function has the same spin alignment as the proton spin in terms of Bjorken scaling variable $x_{bj} \to 1$.
Thus one only needs the $S$-state to identify the nucleon at $z=0$ in the AdS/CFT description. The $P$-wave component of the eigensolution automatically develops as one evolves the eigenstate to $z > 0.$  
This is in close analogy to the solutions of the Dirac-Coulomb equation in QED. The ground-state eigensolution has both S and P waves (upper and lower components). However, only the S wave appears at the origin ($r=0$).

As another example, consider the production of a  $\rho$  meson in $e^+ e^- \to \rho \to \pi^+ \pi^-$ in an electron-positron collider. The vector meson is initially created at a point $x^2=0$  in space-time as an S-wave, and it then propagates to the physical  $\rho$ eigenstate.   The only relevant interpolating operators for a vector meson are 
$\epsilon_\mu\bar \psi \gamma^\mu \psi$ and
$\epsilon_\mu   \bar \psi\sigma ^{\mu  \nu}  \partial_\nu \psi$  where $i \partial_\mu$ is the relative momentum $q_\mu$ in momentum space.
Only the leading-twist interpolating operator, corresponding to the  $\bar \psi(0)\gamma^\mu \psi(0)$  vector current, survives at small interquark separation $x^2 \to 0$;  this is the $z \to 0$ domain where one applies the AdS  boundary condition.  This is also seen immediately from the definition of the electromagnetic current of the quarks in QCD in the interaction picture. 
The Dirac current is consistent with chiral symmetry.
The pseudotensor enters via the Pauli interpolating operator, and it has higher twist.  
The higher orbital components which are present in the full eigenstate do not couple at $x^2=0$ ($z=0$) to the local current.  Thus the decay of a vector meson to lepton pairs or its production in electron-positron annihilation proceeds only through the S wave; i.e., the Dirac current.   In general, one can use the  OPE to determine the contribution of interpolating operators at short distances from the operator's twist~\cite{Brodsky:1980ny}.    The leading twist operators have minimal $L$ and dominate at short distances, but this does not imply that the holographic or LF descriptions require fixed $L$. No symmetries are violated.

In summary, the dominance of components with minimal $L$ at short distances $x^2 \to 0$ or $z \to 0$  is required by the OPE, and is not in conflict with the underlying chiral symmetry of the full eigenstates.  The dominance of the minimal $L$ component at $z \to 0$  also yields the dimensional counting rules for form factors at high $Q^2$, as shown by Polchinski and 
Strassler~\cite{Polchinski:2001tt}. In the case of mesons, the interpolating operators that we use at $z=0$ give the correct definition of decay constants, such as $f_\rho$ and $f_\pi.$  A similar OPE analysis is used to analyze the evolution of hadronic distribution amplitudes.    Analogous local operators are used in lattice gauge theory to create and identify specific hadrons at the lattice boundary.

It is also important to note that the relativistic light-front equation and its eigensolutions can be derived  starting from the light-front Hamiltonian formalism, independent of AdS space considerations~\cite{deTeramond:2008ht}.
Eigenstates in the LF formalism are in general mixtures of components with different $L$, whereas only the component with minimum twist, and thus only the minimum $L$ projection appears at short distances. Of course, it is extremely interesting that a dual 5-dimensional background description emerges from LF QCD, and that a geometrical representation of the results can be established; thus the powerful geometrical methods from string theory can be used in the description of strongly coupled QCD.

We thank Leonid Ya. Glozman for discussions. This research was supported by the Department of Energy contract
DE--AC02--76SF00515. SLAC-PUB-13422.


\begin{thebibliography}{0}

 \bibitem{deTeramond:2008ht}
  G.~F.~de Teramond and S.~J.~Brodsky,
  Phys.\ Rev.\ Lett.\  {\bf 102}, 081601 (2009).
  
  \bibitem{Klempt:2007cp}
  E.~Klempt and A.~Zaitsev,
  Phys.\ Rept.\  {\bf 454}, 1 (2007).
  
  \bibitem{Glozman:2009fj}
  L.~Y.~Glozman,
  arXiv:0903.3923 [hep-ph].
  
    \bibitem{Brodsky:2008pg}
  S.~J.~Brodsky and G.~F.~de Teramond,
  arXiv:0802.0514 [hep-ph],
  Proceedings of the International School of Subnuclear Physics: Searching for the ``Totally Unexpected" in the LHC Era, Erice, Sicily, Italy, 29 Aug - 7 Sep 2007.
  
 \bibitem{Brodsky:1980ny}
 See for example: 
  S.~J.~Brodsky, Y.~Frishman, G.~P.~Lepage and C.~T.~Sachrajda,
  Phys.\ Lett.\  B {\bf 91}, 239 (1980), and references therein.
  
  \bibitem{Polchinski:2001tt}
  J.~Polchinski and M.~J.~Strassler,
  Phys.\ Rev.\ Lett.\  {\bf 88}, 031601 (2002).

  
\end{thebibliography}
\end{document}